- 1.5 year intercomparison of TLD fading properties was performed
- TL intensity was found to be stable within 10% for all patterns of irradiation
- No difference between gamma-ray an neutron exposure was found
- No significant difference between +18°C and -17°C storage temperature was found

# Thermoluminescence fading studies: Implications for long-duration space measurements in Low Earth Orbit

P. Bilski[1], T. Berger[2], M. Hajek[3], A. Twardak[1], C. Koerner[2], G. Reitz[2]

[1] Institute of Nuclear Physics, Polish Academy of Sciences, Kraków, Poland
[2] German Aerospace Center, Institute of Aerospace Medicine, Cologne, Germany
[3] Institute of Atomic and Subatomic Physics, Vienna University of Technology, Vienna, Austria

Abstract

Within a 1.5 year comprehensive fading experiment several batches of LiF:Mg,Ti and LiF:Mg,Cu,P thermoluminescence detectors (TLDs) were studied. The TLDs originated from two manufacturers and were processed by three laboratories using different annealing and readout conditions. The TLDs were irradiated with two radiation modalities (gamma-rays and thermal neutrons) and were stored at two temperatures (–17.4°C and +18.5°C). The goal of the experiment was to verify the stability of TLDs in the context of their application in long-term measurements in space.

The results revealed that the response of all TLDs is stable within 10% for the studied temperature range. No influence of the radiation type was found. These results indicate that for the properly oven-annealed LiF TLDs, fading is not a significant problem, even for measuring periods longer than a year.

## 1. Introduction

The response of thermoluminescent detectors (TLDs) may exhibit some changes during their storage, both before and after irradiation (Budzanowski, Saez-Vergara et al. 1998; Burgkhardt and Piesch 1983; Driscoll et al. 1985; Horowitz 1990). At each (even low) temperature there is a probability that charge carriers escape from the trapping centres within a TLD, resulting in the so-called "fading" of the TL signal. Also during the time between annealing and exposure, the defect structures, acting as trapping and recombination centres, may undergo some transformations, leading to changes of sensitivity (decrease as well as increase). The main external factors affecting fading are temperature and time of the storage.

In case of space-borne application of TLDs, time is the main element making a difference comparing to ground usage in the context of fading. In radiation protection applications on Earth the typical measuring periods are up to 3 months. Longer exposure durations, in these and other typical applications, are rare. However, in case of measurements of cosmic radiation in Low Earth Orbit (LEO), the situation is quite different and exposure times are commonly ranging from several months up to years. This was especially true for the MATROSHKA (MTR) experiment (Reitz and Berger 2006; Reitz et al. 2009) – the biggest application of TLDs in space, consisting of the exposure of a human phantom equipped with several thousands of TLDs inside and outside the International Space Station (ISS). Within this experiment four measuring periods lasting up to 1.5 years were performed in the years 2004 to 2011 (Berger et al. 2012). The specificity of organization of space experiments extends this time by additional several months required between the preparation of the detectors in a laboratory and their delivery to the ISS. The available data on performance of TLDs for such long storage periods are in principle missing.

The temperature conditions in the ISS are not different from that encountered on Earth. The temperature inside the station may be considered as typical room temperature (RT), up to 23°C. During the MTR measurements outside the ISS temperature sensors were located inside the phantom and the measured values fluctuated between –20°C and +25°C.

The long duration of the measurements raised the question whether the results of the MTR experiment were not significantly biased by fading effects. In order to answer this question, three of the laboratories participating in the MTR project (IFJ Krakow, DLR Cologne, ATI Vienna) have realized an inter-comparison experiment aimed on the long-term studies of TLD response stability.

## 2. Materials and methods

### 2.1 TL detectors

The investigations applied the actual batches of LiF TLDs, which were previously used in space within the MTR experiment comprising six batches of LiF:Mg,Ti and one of LiF:Mg,Cu,P detectors. In case of DLR and ATI these were TLD-700 and TLD-600 (LiF:Mg,Ti) chips manufactured by Thermo Fisher Scientific Inc., while IFJ used the own production MTS-7 and MTS-6 (LiF:Mg,Ti) as well as MCP-7 (LiF:Mg,Cu,P) pellets. Each laboratory prepared, processed and evaluated the TLDs using their own standard procedures (identical with those used within the MTR experiment), which were different in each case. The details of the used procedures are summarized in the Table 1.

→ *Table 1*

### 2.2 Experimental procedure

The fading study was accomplished at IFJ Krakow. All TLDs, after annealing at their home laboratories, were sent to IFJ and were divided into several groups, which then underwent different irradiation and storage procedures. Two storage locations – a laboratory room (nominal RT, the real average of 18.5°C) and a freezer (–17.4°C) – and two radiation modalities – gamma rays and thermal neutrons – were used. The use of thermal neutrons was intended for checking a possible influence of ionization density on fading effects, as densely ionizing radiation is abundant in space. Thermal neutron irradiations were performed with a polyethylene-moderated Pu-Be source and were applied only to TLDs highly enriched in $^6$Li isotope: TLD-600 and MTS-6. Gamma exposures were realized using the $^{137}$Cs source at the IFJ reference laboratory. The temporal irradiation pattern consisted of exposures at the beginning, in the middle and close to the end of the storage at IFJ (see Table 2). The gamma-ray dose was always 40 mGy in terms of absorbed dose in water. In case of thermal neutrons the exact dose was unknown, but the rough estimate of the fluence is $2\times10^7$ cm$^{-2}$. Additionally, for gamma rays, a fourth exposure pattern was applied, with the intention of simulating a continuous irradiation consisting of six dose fractions of 4 mGy each, equally distributed over the storage period, which amounted to the total dose of 24 mGy. The storage period at IFJ lasted 448 days. Afterwards, the TLDs were distributed to the laboratories and read out about 2.5 months later. The total time of the experiment was 560 days.

For each TLD the absorbed dose was determined using evaluation and calibration methods applied normally in the respective laboratories for evaluation of TLDs exposed to unknown doses during space experiments. To check possible differences in calibration applied at ATI, DLR and IFJ, a separate gamma-ray inter-calibration experiment was additionally realized. In the frame of this experiment, sets of TLDs were exposed to the same calibration dose at the three reference calibration laboratories (Laboratory for Calibration of Dosimetric Instruments at IFJ Kraków, Poland; Materialprüfungsamt Nordrhein-Westfalen in Dortmund, Germany; Department of Radiotherapy of the Medical University of Vienna, Austria) and evaluated together. The results confirmed an agreement of calibrations between the reference laboratories within 2.5%.

→ *Table 2*

## 3. Results

Figures 1 and 2 show examples of the measured glow-curves for various irradiation patterns and different storage temperatures. It is apparent that the storage and irradiation conditions had rather small influence on the glow-curve shape. Glow-curves measured at DLR (Figs 1a, 2a) exhibit somewhat increased low-temperature tails, compared to ATI (Figs 1b, 2b) and IFJ (Figs 1c, 2c) glow-curves. The reason for this behaviour lies in the application of a pre-heat cycle before readout at the latter laboratories. On the other hand, the main peak height of DLR TLDs seems to be very stable (Figs 1a, 2a). Glow-curves of the samples stored in a freezer (Fig. 2) are little increased compared to RT storage (Fig. 1), in particular on the low-temperature side of the main peak. Glow-curves of MTS-7 (Fig. 1c) and MTS-6 (not shown) detectors irradiated at the beginning of the storage show stronger presence of peak 4 (at ~ 200ºC), indicating some evolution of the structure of corresponding trapping sites during the first days after annealing (time between annealing at the first irradiation was shorter in case of IFJ detectors than for ATI and DLR). For ATI TLD-700 a small decrease of the main peak height during storage was observed (Fig. 1b). For MCP-7 no changes in glow-curve shape were present.

→ *Fig. 1 (one column)*

→ *Fig. 2 (one column)*

Figures 3 and 4 illustrate how the observed small changes in the glow-curve shape influence the performance of TLDs, i.e., how they affect the measured doses. The results are presented as the relative response. For gamma-ray exposures (Fig.3) the relative response is defined as the ratio of the measured dose to the true dose value. For neutron exposures (Fig. 4), where the true dose value was unknown, the measured doses were normalized to the values obtained for the exposure #3. The results for gamma rays reveal that nearly all data points lay within 10% around unity (three data points are distanced by exactly 11%). Moreover, over 80% of the data points were found within 5% around unity. The results for the freezer storage seems to be somewhat more scattered than those for the RT storage, but there is no direct trend. Similar results were obtained also for the neutron exposures as given in Figure 4. The higher uncertainties are caused probably by a not perfect uniformity of the neutron radiation field. This lack of a significant fading is contrary to some recent papers (Carinou et al. 2011; Gilvin 2007), which report fading up to 20% within 6 months. This may be explained by the simplified reader annealing procedures used in those studies, instead of the full oven annealing as applied in the present work.

→ *Fig. 3 (two columns!)*

→ *Fig. 4 (one column)*

## 4. Conclusions

The obtained results indicate a high stability of the studied thermoluminescent detectors disregarding their producers and the applied annealing and readout conditions. Nearly all measured doses were within 10% from the true values, with over 80% of the results within 5%. Even the longest storage period (about 1.5 years) did not influence the results in a significant way. No difference was observed between gamma and neutron exposures, as well as between the two studied storage temperatures. These results indicate that for the properly oven-annealed LiF TLDs, the fading is not a significant problem, even for measuring periods of more than a year.

Acknowledgments
This work was partly funded by the European Commission in the frame of the FP7 HAMLET project (Project #218817) and by the Polish National Science Center (project DEC-2011/01/B/ST2/02450).

References

Berger, T., Bilski, P., Hajek, M., Puchalska, M., and Reitz, G., 2012. The MATROSHKA Experiment: Results and Comparison from EVA (MTR-1) and IVA (MTR-2A/2B) Exposure. Radiat.Res.(submitted).

Budzanowski, M., Saez-Vergara, J.C., Gomez Ros, J.M., Romero, A.M., and Ryba, E., 1998. The Fading of Different Peaks in LiF:Mg, Cu, P (MCP-N and GR-200A) TL Detectors. Radiat.Meas. 29, 361-364.

Burgkhardt, B. and Piesch, E., 1983. Fading Characteristics of LiF and $Li_2B_4O_7$ TLD Systems Dependent on the Ambient Temperature, the Monitoring Period and the Interval Between Irradiation and Readout. Radiat.Prot.Dosim. 6, 338-340.

Carinou, E., Askounis, P., Dimitropoulou, F., Kiranos, G. et al, 2011. Pre- and post-irradiation fading effect for LiF:Mg,Ti and LiF:Mg,Cu,P materials used in routine monitoring. Radiat.Prot.Dosim. 144, 207-210.

Driscoll, C.M.H., McWhan, A.F., and Richards, D.J., 1985. A comparative study of the sensitivity and fading characteristics of thermoluminescent LiF chips. Radiat.Prot.Dosim. 11, 119-121.

Gilvin, P.J., 2007. Comparison of time effects, decision limit and residual signal in Harshaw LiF:Mg,Ti and LiF:Mg,Cu,P. Radiat.Prot.Dosim. 125, 233-236.

Horowitz, Y.S., 1990. Fading in LiF:Mg,Ti. Radiat.Prot.Dosim. 32, 147-148.

Reitz, G. and Berger, T., 2006. The MATROSHKA facility - Dose determination during an EVA. Radiat.Prot.Dosim. 120, 442-445.

Reitz, G., Berger, T., Bilski, P., Facius, R. et al, 2009. Astronaut's organ doses inferred from measurements in a human phantom outside the international space station. Radiat.Res. 171, 225-235.

Table Captions

Table 1. Parameters of the measurement procedures used by ATI, DLR and IFJ.

Table 2. Data of TLD storage before and after exposures. The given values are averages over dates scattered within a few days (small differences in annealing and readout dates at the laboratories).

Figure Captions

Fig.1. Comparison of glow-curves of $^7$LiF:Mg,Ti detectors (TLD-700, MTS-7) irradiated at different times (storage at RT).

Fig.2. Comparison of glow-curves of $^7$LiF:Mg,Ti detectors (TLD-700, MTS-7) stored at different temperatures (irradiation performed at the beginning of the storage).

Fig. 3. Relative response (dose measured divided by the dose delivered) of TLDs after gamma-ray exposure. Shape of a symbol indicates the number of the exposure. For each exposure, the order of plotting from left to right is the following: DLR; ATI; IFJ integral; IFJ peak height.

Fig. 4. Relative response (normalized to the exposure #3 at the end of the storage period) of $^6$LiF:Mg,Ti TLDs (TLD-600, MTS-6) after thermal neutron exposure.

Table 1

| Parameter | ATI | DLR | IFJ |
|---|---|---|---|
| TL reader | TL-DAT.II | Harshaw 5500 | RA'94 |
| Heating method | Contact | Hot nitrogen gas | Contact |
| Neutral gas flow | Nitrogen | Nitrogen | Argon |
| Heating rate | 5 °C/s | 5 °C/s | 10 °C/s |
| Pre-heat | 120 °C (30 min) | no preheat | 120 °C (30 min) |
| Annealing cycle: | | | |
|   LiF:Mg,Ti | 400°C (1 h) slow cooling* | 400°C (1 h)+100°C (2h) slow cooling* | 400°C (1 h)+100°C (2h) fast cooling* |
|   LiF:Mg,Cu,P | - | - | 240°C (10 min) fast cooling* |
| Evaluation method | Peak height | Peak height | Peak height and peak integral |

(*) – "slow cooling" means cooling of TLDs inside an oven, "fast cooling" means removing of TLDs from the hot oven after the end of the annealing period; because of different ovens, cooling rate for DLR was faster than for ATI.

Table 2

| Irradiation | Days between annealing and irradiation | Days between irradiation and readout |
|---|---|---|
| 1 – "start" | 10 | 550 |
| 2 – "middle" | 240 | 320 |
| 3 – "end" | 460 | 100 |

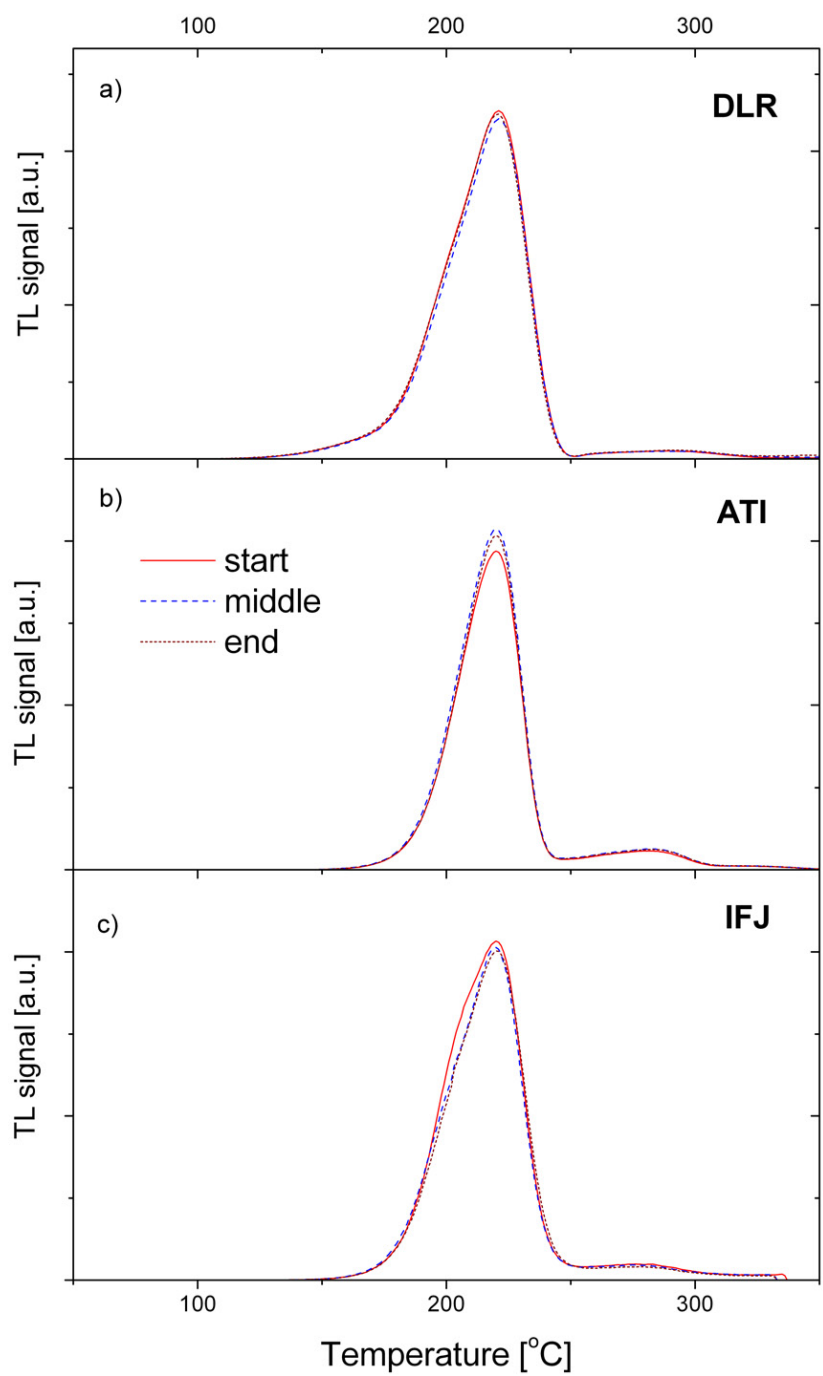

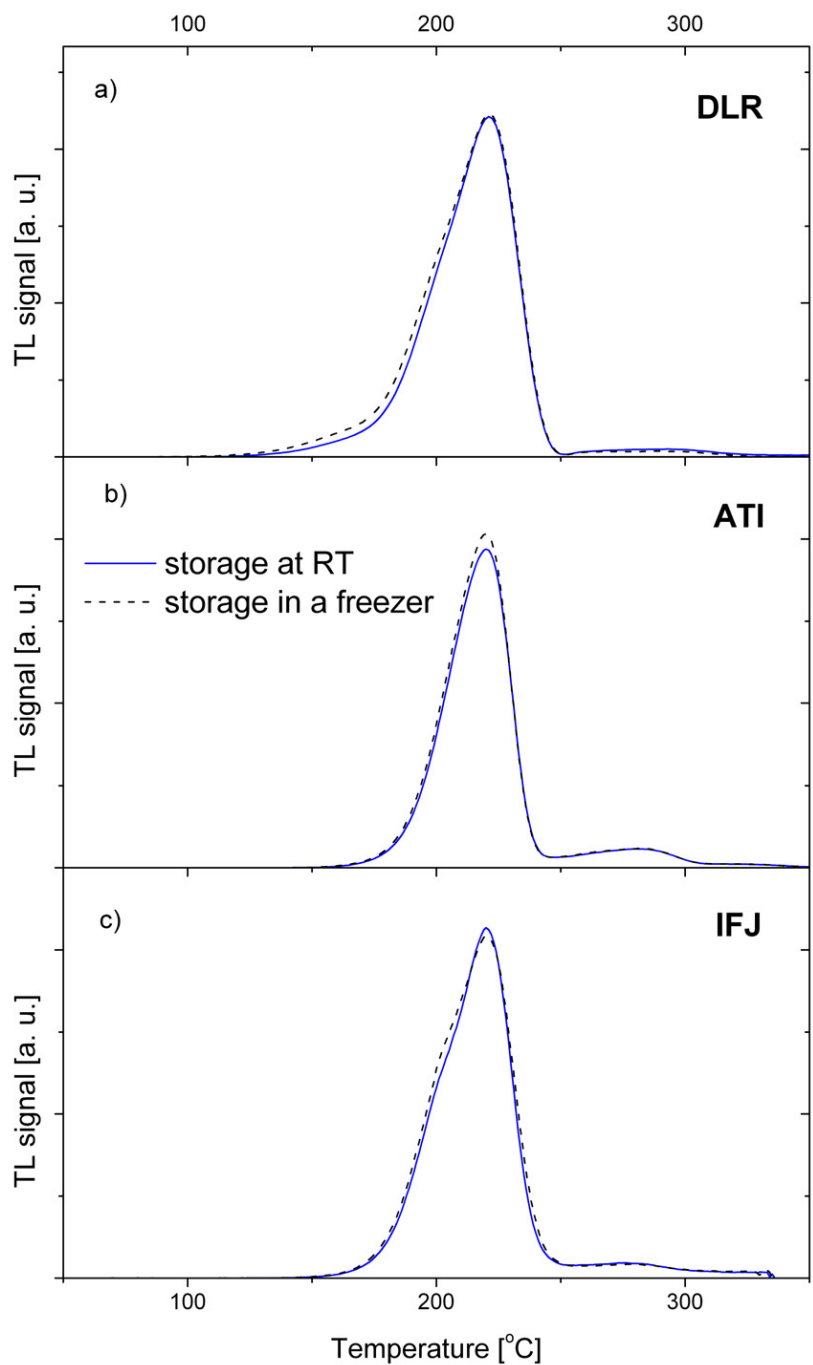

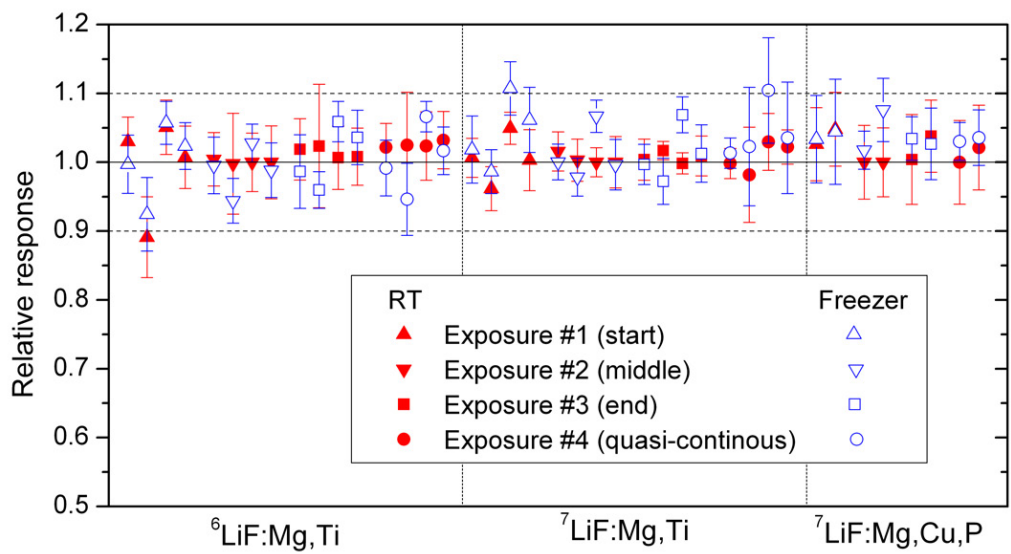

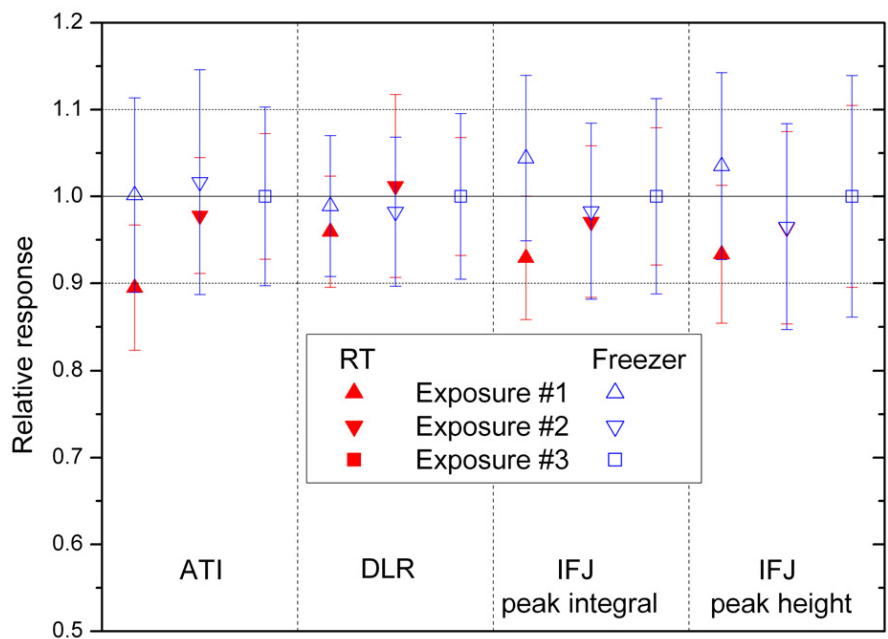